\documentclass[a4paper]{article}
\usepackage{url}
\usepackage{cite}
\usepackage{xcolor}

\usepackage{INTERSPEECH2022}

\title{Efficient Training of Audio Transformers with Patchout}

\name{Khaled Koutini$^{1,2}$, Jan Schlüter$^1$, Hamid Eghbal-zadeh$^{1,2}$, Gerhard Widmer$^{1,2}$}
\address{Institute of Computational Perception$^1$ \& LIT AI Lab$^2$, Johannes Kepler University Linz, Austria}
\email{first.last@jku.at}

\begin{document}

\maketitle
\begin{abstract}
The great success of transformer-based models in natural language processing (NLP) has led to various attempts at adapting these architectures to other domains such as vision and audio.
Recent work has shown that transformers can outperform Convolutional Neural Networks (CNNs) on vision and audio tasks. However, one of the main shortcomings of transformer models, compared to the well-established CNNs, is the computational complexity. In transformers, the compute and memory complexity is known to grow \emph{quadratically} with the input length. Therefore, there has been extensive work on optimizing transformers, but often at the cost of degrading predictive performance.
In this work, we propose a novel method to optimize and regularize transformers on audio spectrograms.
Our proposed models achieve a new state-of-the-art performance on Audioset and can be trained on a single consumer-grade GPU. Furthermore, we propose a transformer model that outperforms CNNs in terms of both performance and training speed.\footnote{Source code and pretrained models: \url{https://github.com/kkoutini/PaSST}}
\end{abstract}

\section{Introduction}
\label{sec:intro}

The transformer architecture~\cite{vaswani2017attention} has proven very successful in sequence modeling. It allows learning dependencies between different items in the sequence regardless of their positions or their separation in the sequence. Transformers are the state-of-the-art models in different natural language processing tasks~\cite{devlinCLT19bert,so21Primer,YamadaASTM20luke}. More recently, they have been adapted to computer vision~\cite{dosovitskiyB0WZ21VIT,Liu21swine,ZhuSLLWD21detr} by extracting small patches from the input image and adding a learnable positional encoding to each patch. The resulting patches form a sequence that can be fed to the transformer. These vision transformer models achieve state-of-the-art performance on image classification tasks, but require large amounts of training data (e.g, in Vision Transformer (ViT)~\cite{dosovitskiyB0WZ21VIT}), or heavily depend on extensive data augmentation and knowledge distillation from a CNN model (e.g, in Data-efficient Image Transformers (DeiT)~\cite{TouvronCDMSJ21deit}). Gong et al.~\cite{gong21ast} further adapted vision transformers to audio spectrograms, achieving state-of-the-art performance on Audioset~\cite{audioset2017Gemmeke} by using pre-trained models from computer vision and using overlapping patches from audio spectrograms for fine-tuning.

\begin{figure}[tb]
\centering
\centerline{\includegraphics[width=8.5cm]{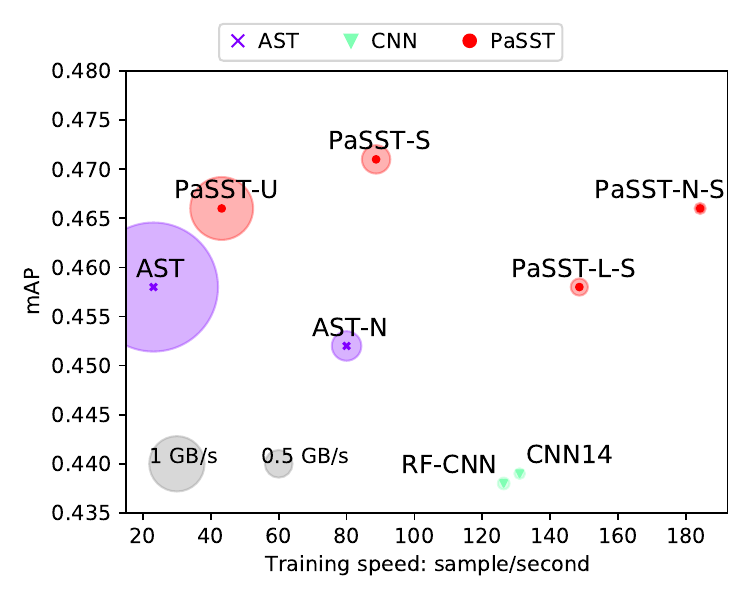}}
\caption{Performance vs training speed on Audioset. The radius of the circle indicates the required GPU memory per sample for training. On the largest publicly available audio dataset, our approach PaSST-S can reach the state-of-the-art performance in less than 2 days on a single consumer GPU. Details are presented in Table~\ref{tab:audioset:results}.}

\label{fig:res}
\end{figure}

The transformer architecture consists of a series of self-attention layers~\cite{vaswani2017attention}. Each layer relies on calculating a distance between each pair of items from the input sequence. Although this allows each input item to attend to any other item in the sequence, this results in a complexity of $\mathcal{O}(n^2)$ with respect to the input sequence length $n$, in terms of both memory and computation effort. Reducing the quadratic complexity has been the target of several approaches in natural language processing, the idea being to restrict each input item (token) to attend only to a pre-selected subset of input items (tokens). One example is to allow attending only to neighbours inside a sliding window~\cite{WangNMNX19MultipassageBert,SukhbaatarGBJ19adaptivespan}. Additionally, Kitaev et al.~\cite{KitaevKL20reformer} use locality-sensitive hashing to approximate attention, reducing the attention complexity to $\mathcal{O}(n\log{}n)$. BigBird~\cite{ZaheerGDAAOPRWY20bigbird} combines sliding windows, global attention, and random interaction between the sequence items. Masking portions of the input sequence has been shown to be an extremely effective method for training encoder/decoder transformers in NLP~\cite{devlinCLT19bert}. In computer vision, the idea of removing patches during \emph{inference} was investigated to assess the vision transformer's \emph{robustness} against input perturbations~\cite{naseer2021Intriguing}. 

In this paper, we focus on applying transformers to \emph{audio processing}. We address the shortcomings of current audio transformers from the aspect of computational complexity and memory requirements by introducing a new simple yet effective method for training transformers on spectrograms.
In summary, the main contributions of our work are as follows:
\begin{itemize}
    \item  We propose \emph{Patchout}, a method that significantly reduces the computation and memory complexity of training transformers for the audio domain. Patchout also functions as a regularizer, improving the generalization of the trained transformers.
    \item We disentangle the transformer's positional encoding~\cite{dosovitskiyB0WZ21VIT,TouvronCDMSJ21deit,gong21ast} into time and frequency positional encoding, allowing for straightforward inference on audio snippets of variable length without the need for fine-tuning or interpolating positional encodings.
    \item We investigate additional methods for reducing training complexity and demonstrate how they affect performance on the larger general-purpose Audioset as well as domain-specific downstream tasks.
\end{itemize}
Our proposed models can achieve state-of-the-art performance on several audio tagging and classification tasks using a single consumer GPU, in a relatively short time (see Figure~\ref{fig:res}). When different complexity reduction methods are combined, the models outperform CNNs in terms of training speed and memory requirements, in addition to generalization.

\section{The Patchout faSt Spectrogram  Transformer {(P\MakeLowercase{a}SST)}}
\label{sec:arch}

The Vision Transformer (ViT)~\cite{dosovitskiyB0WZ21VIT} works by extracting small \emph{patches} from an input image and projecting these patches linearly onto a sequence of embeddings. The sequence is then augmented by adding trainable positional encodings as biases to the input sequence. A special classification embedding (classification token) is then appended to the sequence, which is connected to a classifier after the self-attention layers. In  Data-efficient Image Transformers (Deit)~\cite{TouvronCDMSJ21deit} another special embedding for distillation  (distillation token) is added. 
Gong et al.~\cite{gong21ast} showed that \emph{overlapping} the extracted patches improves the performance when training ViT on spectrograms. On the other hand, patch overlapping increases the total number of patches, i.e., the input sequence length. Therefore, overlapping greatly increase the memory and compute requirements for training the transformers. We propose a new method called \emph{Patchout} (Section~\ref{sec:patchdrop}) to overcome these issues. 

\begin{figure}[htb]
 \centerline{\includegraphics[width=8.5cm]{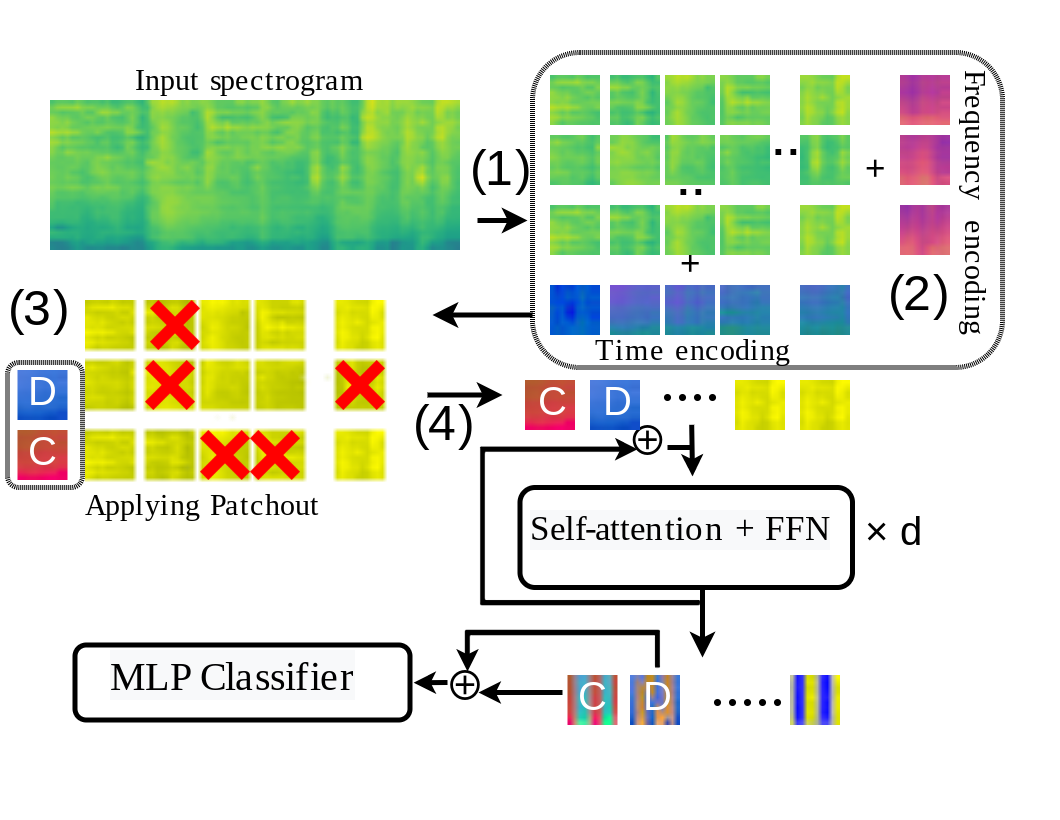}}
\caption{The Patchout transformer (PaSST) architecture as explained in Section~\ref{sec:arch}. The Self-attention layer + FFN (Feed-forward network) is explained in detail in~\cite{dosovitskiyB0WZ21VIT}. C: classification token. D: distillation token (only for models based on DeiT). }
\label{fig:passt:arch}
\end{figure}

Figure~\ref{fig:passt:arch} summarizes the proposed transformer architecture: The pipeline starts at the upper left with an audio spectrogram being fed into the model as input. (1) is the patch extraction and linear projection steps as explained in~\cite{dosovitskiyB0WZ21VIT}. In (2), frequency and time positional encodings are added, as discussed below. In (3), we apply \emph{Patchout} as explained in Section~\ref{sec:patchdrop}, and add the classification token. In (4), we flatten the sequence and pass through $d$ layers blocks of self-attention ($d$ is the depth of the transformer). Finally, a classifier operates on the mean of the transformed C and D tokens.

In Addition to  \emph{Patchout} (step 3 in Figure~\ref{fig:passt:arch}), the second main difference between our work and previous work~\cite{dosovitskiyB0WZ21VIT,TouvronCDMSJ21deit,gong21ast} is that we disentangle the positional encoding for the time and frequency dimensions, and as a result, we have two positional encodings: one representing the frequency, and one for time (step 2 in Figure~\ref{fig:passt:arch}). This makes inference and the tuning of the pre-trained models on downstream tasks with shorter audio length simpler.
When fine-tuning or inference on shorter audio clips, we simply crop the time positional encoding parameters, without changing the frequency encoding parameter.

\subsection{Complexity Analysis}
\label{sub:sec:complexity}
Multi-head attention layers~\cite{vaswani2017attention} rely on computing a distance between each pair of positions in the input sequence (in the form of an attention matrix), therefore having a complexity of $\mathcal{O}(n^2)$ where $n$ is the input sequence length. As the sequence length grows  -- for example, when overlapping input patches or for longer audio clips -- the compute and memory requirements quickly become problematic. More specifically, given an input of $b$ samples of the dimension $ ( n\times e)$, where $b$ is the batch size, $n$ is the input sequence length, $e$ is the embeddings size, each \emph{multi-head} attention layer projects 
each input sample to $h$ query $Q$, key $K$, and value $V$ matrices, where $h$ is the number of attention heads~\cite{vaswani2017attention}.  Each of $Q$, $K$ and $V$ 
has a shape of $ ( n \times \frac{e}{h}) $. The \emph{attention matrix} $A$ is then computed by the matrix multiplication $Q \times K^T$, scaling, and applying the soft-max activation function $A = Softmax(\frac{Q \times K^T}{\sqrt{d}})$.
$A$  has a shape of $( n \times n)$ and 
is multiplied with $V$ giving the attention output: $A \times V$ resulting in a new sequence with the same shape as the input $ ( n\times e)$.
As a result, the computation complexity (and memory requirements) for all the operations on the attention matrix $A$ grow quadratically $\mathcal{O}(n^2)$ with sequence length $n$, while the operations in the rest of the network have a linear complexity relationship with $n$~\cite{vaswani2017attention,dosovitskiyB0WZ21VIT,TouvronCDMSJ21deit}.
In short, reducing the sequence length would have a large impact on the computational complexity of these models.

\subsection{Patchout}
\label{sec:patchdrop}
Motivated by (a) the impact of reducing the sequence length on the computation complexity of training transformer models; (b) the fact that audio events are expected to be spread out in time and frequency 
in an audio clip; 
(c) the insight that CNNs can benefit from having a small receptive field during training for different audio tasks, as shown in~\cite{Koutini2019Receptive}, we propose \emph{Patchout}, a method to efficiently train transformer models on audio spectrograms. The idea is to drop parts of the transformer's input sequence when training, encouraging the transformer to perform the classification using an incomplete sequence. We first extract small overlapping patches from the input spectrograms and linearly project them to vectors, forming the transformer input sequence. We augment the patches with both frequency and time encoding. When training, we randomly drop parts of the sequence, reducing the sequence length, and effectively regularizing the training process. Similar to DropOut~\cite{JMLR:v15:srivastava14a:dropout}, during inference, the whole input sequence is presented to the transformer.  We distinguish between different types of Patchout as follows: 

\textbf{Unstructured Patchout} is the basic form of Patchout, where we select the patches randomly regardless of their position. We refer to models trained with this method as \emph{PaSST-U}.

\textbf{Structured Patchout:} We randomly pick some \emph{frequency bins}/\emph{time frames} and remove a whole column/row of extracted patches. This structure is inspired by SpecAugment~\cite{ParkCZCZCL19Specaugment}. We refer to models trained with this method as \emph{PaSST-S}.



\subsection{Further Complexity Reduction Methods}
\label{sub:sec:further:complex:reduce}

\subsubsection{Reducing the extracted patches' overlap}
Reducing the overlap between patches results in a lower number of extracted patches, and therefore a smaller  transformer input sequence length. However, Gong et al.~\cite{gong21ast} showed that reducing the overlap (
or training without overlapping) degrades the performance of the transformer on Audioset. Patchout can also be used even when there is no overlap between patches. We refer to the system without patch overlapping as \emph{PaSST-N}.

\subsubsection{Reducing the depth of the transformer}
\label{sub:sec:reduce:depth}
The depth of the transformer is the number of successive self-attention blocks ( \emph{d} in Figure~\ref{fig:passt:arch}). The depth has a linear relationship with the overall training and inference complexity and influences the total number of parameters of the model. Since we are starting the training from models pre-trained on Imagenet~\cite{dengImagenetLargescaleHierarchical2009} (as explained in Section~\ref{sec:imageNet}), we remove every other self-attention block. This allows us to benefit from the pre-training, compared to removing consecutive blocks, since the residual activations will have a less sudden change. We will refer to the model with the removed blocks as \emph{PaSST-L}. It has $d=7$ self-attention blocks and 50M parameters compared to 87M in the full model.

\section{Experiment Setup}
\label{sec:setup}

 We train our models on Audioset~\cite{audioset2017Gemmeke}, the largest publicly available audio dataset, consisting of around 2 million audio clips from Youtube. The task is to tag the audio clips with labels from 527 possible classes. Furthermore, we fine-tune the models trained on Audioset on various audio classification and tagging tasks, namely, instrument recognition, environmental audio classification, and acoustic scene classification.  

\subsection{Preprocessing and Training Setup}
\label{sec:preprocessing}
We use mono audio with a sampling rate of $32$\,kHz. We extract Mel features from a window of $25$\,ms with a hop length of $10$\,ms, resulting in $128$ mel bands, similar to~\cite{gong21ast}. Kong et al.~\cite{KongCIWWP20panns} showed the importance of balancing Audioset; therefore, we balance our training data using importance sampling. We assign a sampling weight to each sample proportional to the inverse frequency of its label $\frac{1}{\mathit{freq(label)}+100}$. We train on $1,893,693$ (approx.\ 2M) training segments, and evaluate on $18,951$ audio clips. For each epoch, we sample 200k samples from the full 2M Audioset without replacement.
We use the AdamW~\cite{LoshchilovH19adamw} optimizer with weight decay of $10^{-4}$, with a maximum learning rate of $10^{-5}$. We use a linear learning rate decay from epoch 50 to 100, dropping the learning rate to $10^{-7}$ and fine-tune the model for a further 20 epochs.

\subsection{ImageNet Pretraining}
\label{sec:imageNet}

Gong et al.~\cite{gong21ast} showed that using pre-trained models on Imagenet significantly improves their performance on Audioset. Therefore, we will use pre-tranined models in all our experiments.
Our base model is DeiT B↑384~\cite{TouvronCDMSJ21deit}. 
We also achieve a comparable performance using computationally more complex ViT models such as stripped-down ViT-hug224~\cite{dosovitskiyB0WZ21VIT}; by removing half of the self-attention blocks, its depth was reduced to only 16 blocks (with the methods explained in Section \ref{sub:sec:reduce:depth}); this will not be further explored in this paper.
\subsection{Data Augmentation}
\label{sec:augmentation}
The transformer models are very prone to overfitting, therefore data augmentation plays an essential role in the training process \cite{TouvronCDMSJ21deit}.
In our experiments, the following augmentation strategies are used:

 \textbf{Two-level Mix-Up:} We use Mix-up~\cite{zhangMixupEmpiricalRisk2017} since it has been shown to improve performance~\cite{koutini21journal,gong21ast}. We mix both the raw waveforms randomly from the dataset as well as the final spectrograms.

\textbf{Specaugment:} We use SpecAugment~\cite{ParkCZCZCL19Specaugment} by masking up to $48$ frequency bins and $192$ time frames similar to~\cite{gong21ast}.

 \textbf{Rolling:} We roll the waveforms randomly over time.

 \textbf{Random Gain:} We multiply the audio waveforms to change the gain by $\pm$7\,dB.

\section{Results}
\label{sec:results}

\subsection{Audio Tagging on Audioset}

\begin{table}[bt]
\centering
\begin{tabular}{l|l|ll|l|l}
     & mAP & Speed & ($\times$ AST) & Mem & Seq \\ \hline
Baseline~\cite{audioset2017Gemmeke} & .314   &     -   &     &   - & -\\ 
PANNs~\cite{KongCIWWP20panns} &  .439  &     131.0  &  ($5.7\times$)   &   .213 & -  \\ 
AST~\cite{gong21ast}  & .459   &        23.1   & ($1\times$)    &   2.33 & 1212   \\ \hline 
AST~\cite{gong21ast} $\star$ & .459   &    23.1   &  ($1\times$)     & 2.33  & 1190    \\ 
AST-N~\cite{gong21ast} $\star$ & .454   &    80.   &  ($3.4\times$)     & .534 &  498  \\ 
CNN~\cite{koutini21journal} $\star$ & .438    &   126.3  &  ($5.5\times$)    &  .213 &  - \\ \hline  \hline
PaSST-B $\star$ &  .462  &     23.1   &  ($1\times$)    &  2.33&   1190 \\
PaSST-U $\star$ &   .466  &    43.2   &   ($1.9\times$)      &   1.14 &  790  \\ 
PaSST-S    $\star$ &\textbf{ .471} &     88.7  &   ($3.8\times$)   &  .513  &  474\\
PaSST-S-L $\star$ &   .459  &  148.6         & ($6.4\times$) & .311 & 474   \\ 
PaSST-S-N $\star$ &  .466  &      \textbf{184.2}   &   ($8\times$)  &  \textbf{.202} & 254  \\
\hline 
\end{tabular}
\caption{Single-model results on Audioset. $\star$ indicates our run. mAP is the mean average precision (also referred to as precision/recall area under curve).  Speed: training throughput in spectrograms per second on an Nvidia Titan RTX GPU (showing the speedup compared to AST~\cite{gong21ast}). Mem: the required GPU memory to train per sample. Seq: the training sequence length.  \textbf{B}: Baseline without Patchout. \textbf{U}: Unstructured Patchout. \textbf{S}: Structured Patchout. \textbf{N}: no-overlap of the extracted patches. \textbf{L}: lighter model with reduced depth=7. }
\label{tab:audioset:results}
\end{table}

Table~\ref{tab:audioset:results} shows the mean average precision mAP 
(also referred to as precision-recall area under-curve) 
results on Audioset~\cite{audioset2017Gemmeke}. As can be seen, the proposed model \emph{PaSST} achieves a new state-of-the-art performance on the largest available audio tagging dataset. The proposed model outperforms AST~\cite{gong21ast} and significantly outperforms CNNs. Using Patchout not only improves the performance of the transformer architecture, but also increases the training speed approximately \emph{4 times}, and reduces the required GPU memory to \emph{less than $25\%$}. As a result, it is possible to train PaSST on a single Nvidia RTX 2080ti (consumer GPU), achieving state-of-the-art performance in 50 hours. Furthermore, \emph{PaSST-L-S} (with a scaled down depth of $d=7$) and \emph{PaSST-S-N } (without patch overlap) significantly outperform CNNs while maintaining a higher training throughput, and with similar GPU memory requirements. \emph{PaSST-S-L} and \emph{PaSST-S-N } can be trained on a single GPU to reach $.459$ and $.466$  mAP in approximately 25 hours. 
Applying Patchout on the transformer without overlap (\emph{PaSST-S-N}) outperforms the baseline \emph{PaSST-B} (without Patchout) and \emph{AST}~\cite{gong21ast} while being up to 8 times faster, and requiring less than 10\% of the GPU memory for training. The results are also illustrated in Figure~\ref{fig:res}. The only difference between the baseline \emph{PaSST-B}  and the \emph{AST} model is the positional encoding. \emph{AST}~\cite{gong21ast}, like vision transformers~\cite{dosovitskiyB0WZ21VIT,TouvronCDMSJ21deit}, employs grid positional encoding. \emph{PaSST-B}, on the other hand, utilises disentangled time and frequency positional encoding (see Section~\ref{sec:arch}).

\begin{table}[hbt]
\centering
\begin{tabular}{l|l}
     & mAP  \\ \hline
Baseline~\cite{audioset2017Gemmeke} & .314   \\ \hline
PSLA (Ensemble-S)~\cite{gong_psla}& .469 \\
PSLA (Ensemble-M)~\cite{gong_psla}& .474 \\
AST (Ensemble-M) ~\cite{gong21ast}  & .475       \\  
AST (Ensemble-M) ~\cite{gong21ast}  & .485       \\ \hline 
PaSST-S S16,14 (2 models)$\star$ & .486  \\
PaSST-S S10-16 (4 models)$\star$ & .493  \\
PaSST-S S10-16  (5 models)$\star$  &\textbf{ .495}  \\
PaSST-S S10-16  (9 models)$\star$ &\textbf{ .496}  \\
\hline 
\end{tabular}
\caption{Ensemble results on Audioset. $\star$ indicates our run. mAP is the mean average precision (also referred to as precision/recall area under curve). }
\label{tab:audioset:results_ens}
\end{table}

Table~\ref{tab:audioset:results_ens} shows the result of ensemble models. We ensemble models with different  overlap values between input patches (Figure~\ref{fig:passt:arch}). \textbf{S} indicates the patches stride, S16 means no overlap between the patches. The first ensemble (2 models) averages the logits of a model with no patches overlap and a model with an overlap of 2 (stride=14). S10-S16 indicates that the models used have strides of 10,12,14 and 16.

\subsection{Fine-tuning and Transfer to Downstream Tasks}
\label{sec:downstream:results}
We fine-tune the pre-trained (on Audioset) models on several downstream audio tagging and classification tasks with different dataset sizes, Table~\ref{tab:finetune:results} summarizes the results. \emph{PaSST-(B,U,S)} models use the pre-trained \emph{PaSST-S} on Audioset, but for fine-tuning, we use no Patchout, unstructured Patchout, and structured Patchout respectively. 
 It is worth noting that the transformer models can be fine-tuned using a small number of epochs. The results suggest that researchers and practitioners can use pre-trained PaSST and fine-tune them on downstream tasks without the need for large computational resources. 
 
In summary, fine-tuning the transformer model outperforms state-of-the-art CNNs on all tasks. Patchout results in significant speedups and, in many cases, improved generalization. When combined with Structured Patchout (\emph{-S}), reducing complexity by removing patch overlap (\emph{-N}) performs better than reducing transformer depth (\emph{-L}) and enables faster fine-tuning.

We only replace the MLP classifier in the pre-trained models for fine-tuning. When we use Patchout, we randomly remove roughly half of the input sequence. Each experiment was repeated three times, and the average results are reported. The speedup in  Table~\ref{tab:finetune:results} is relative to  \emph{PaSST-B} and is rounded up to the nearest integer. Details on the setup of each task can be found in our github repository.
\\
\textbf{Polyphonic Musical Instrument Recognition:}
The task here is to detect all the instruments present in an audio clip. The \emph{OpenMIC} dataset \cite{humphrey2018openmic} consists 20,000 audio clips. Each clip is 10 seconds long and can be assigned multiple tags out of 20 classes. The metric for the task is the mean average precision.
The state-of-the-art methods for this task are CNNs with restricted receptive fields~\cite{koutini21journal}. \emph{PaSST-S-N} reaches the state-of-the-art performance in less than 30 minutes on a single consumer GPU.
\\
\textbf{Environmental Sound Classification:}
The \emph{ESC50} dataset \cite{piczak2015dataset} consists of 2,000 environmental 5-second audio clips. The task is to classify each clip into one out of 50 possible classes. We report the accuracy averaged over the 5 official folds~\cite{piczak2015dataset}. All PaSST variants (with Patchout) can be fine-tuned on this dataset in less than 5 minutes on a single GPU. The state-of-the-art performance was achieved using the AST transformer model~\cite{gong21ast}. 
The difference between AST and PaSST-B is in the positional encoding, as explained in Section~\ref{sec:arch}.
\\
\textbf{Acoustic Scene Classification:}
The task is to recognize the acoustic scene of 10-second audio clips. We use the TAU Urban Acoustic Scenes 2020 Mobile dataset~\cite{Heittola2020} as used in the DCASE 2020 challenge (\emph{DCASE20}). The audio clips are recorded with different devices and further simulated devices are introduced. The performance is measured using accuracy on a dataset including  unseen devices. The first place in the challenge used CNNs~\cite{Suh2020task1a}. Patchout accelerates training on this task, reaching state-of-the-art in less than an hour. Patchout also allows for fine-tuning on a single consumer GPU. It does, however, lead to a decrease in accuracy.
\\
\textbf{Sound Event Recognition (Tagging) on FSD50K:}
The \emph{FSD50K} dataset \cite{fonsecaFPFS22FSD50K} consists of 51K audio clips 
annotated with 200 sound event classes taken from the Audioset ontology~\cite{audioset2017Gemmeke}. The dataset contains 100 hours of audio and is the second largest publicly available general purpose sound event recognition dataset after Audioset. Furthermore, the FSD50K evaluation set is of high quality, with each evaluation label being double-checked and assessed by two to five independent annotators~\cite{fonsecaFPFS22FSD50K}. The reported results are on the official evaluation subset of FSD50K using the best model on the validation subset. 
The state-of-the-art in \emph{PSLA}~\cite{gong_psla}  is achieved through CNN architecture and a collection of performance-improving methods such as ImageNet pre-training, label enhancement, balancing, data augmentation, and weight averaging.
On this dataset, our approach significantly outperforms the current state-of-the-art. Fine-tuning \emph{PaSST-S} and \emph{PaSST-S-N} takes less than 2 hours and 1 hour, respectively.

\begin{table}[ht]
\centering
\begin{tabular}{l|l|l|l|l}
     & OpenMIC & ESC50 &  DCASE20 & FSD50K  \\ \hline
Baseline &   .795~\cite{humphrey2018openmic} &     76.9~\cite{piczak2015dataset}  & 54.1~\cite{Heittola2020}& .434~\cite{fonsecaFPFS22FSD50K}  \\ 
SOTA&  .831~\cite{koutini21journal}  &    95.6~\cite{gong21ast}      &   73.7~\cite{Suh2020task1a}  & .558~\cite{gong_psla} \\ \hline
-B    $\star$ &.837$\mid 1\times$ &96.3$\mid 1\times$&\textbf{76.3}$\mid 1\times$& .649$\mid 1\times$\\
-U $\star$ & \textbf{.843}$\mid 4\times$  & 96.5$\mid 2\times$& 75.6$\mid 4\times$&.639$\mid 4\times$  \\ 
-S $\star$ & \textbf{.843}$\mid 4\times$  &\textbf{96.8}$\mid 2\times$&  75.6$\mid 4\times$  & \textbf{.653}$\mid 4\times$ \\ 
-S-L $\star$ & .841$\mid 6\times$ &  95.5$\mid 2\times$&  73.7$\mid 6\times$ & .584$\mid 6\times$ \\ 
-S-N $\star$ & .840$\mid 8\times$  & 96.4$\mid 4\times$&  73.9$\mid 8\times$ & .637$\mid 8\times$ \\ 
\hline 
\end{tabular}
\caption{Results in performance and speedup compared to the base model PaSST-B for the downstream tasks: polyphonic instrument tagging using OpenMIC~\cite{humphrey2018openmic} dataset (mean average precision),  Environmental Sound Classification ESC50~\cite{piczak2015dataset} (accuracy), Cross-device Acoustic Scene Classification DCASE20~\cite{Heittola2020} (accuracy). Sound Event Recognition (Tagging) in FSD50K~\cite{fonsecaFPFS22FSD50K}(mean average precision). The second part of the table compares different PaSST variants ( Table~\ref{tab:audioset:results}).}
\label{tab:finetune:results}
\end{table}

\section{Conclusion}
\label{sec:conclusion}
We propose a new method for efficiently training transformers on audio spectrograms, achieving state-of-the-art performance on Audioset as well as several downstream tasks. Furthermore, Patchout significantly reduces compute complexity and memory requirements for training transformers.
We investigate additional methods for reducing training complexity and propose two models, \emph{PaSST-S-L} and \emph{PaSST-S-N}, that outperform CNNs while having a faster training speed and comparable memory requirements.
Our pre-trained models can be fine-tuned on several audio downstream tasks with little resources and little additional training time.

\section{ACKNOWLEDGMENT}
This work has been supported by the COMET-K2 Center of the Linz Center of Mechatronics (LCM) funded by the Austrian Federal Government and the Federal State of Upper Austria.
The LIT AI Lab is financed by the Federal State of
Upper Austria. The computational results presented have been achieved in part using the Vienna Scientific Cluster (VSC).

\bibliographystyle{IEEEtran}
\bibliography{refs}

\end{document}